\documentclass[a4paper,fleqn,usenatbib]{mnras}

\usepackage{newtxtext,newtxmath}

\usepackage[T1]{fontenc}
\usepackage{ae,aecompl}

\usepackage{graphicx}	
\usepackage{amsmath}	
\usepackage{amssymb}	
\usepackage{siunitx}
\usepackage{makecell}
\usepackage{multirow}
\usepackage[dvipsnames]{xcolor}
\usepackage[bottom]{footmisc}
\usepackage{mathtools}
\usepackage{booktabs} 
\usepackage{siunitx} 
\usepackage{pgfplotstable}
\usepackage{csvsimple}
\usepackage{float}
\usepackage{adjustbox}
\usepackage{color, colortbl}
\usepackage{resizegather}

\DeclarePairedDelimiterX{\infdivx}[2]{(}{)}{%
  #1\;\delimsize\|\;#2%
}

\definecolor{LightCyan}{rgb}{0.88,1,1}

\title[Machine learning haloes]{An interpretable machine learning framework for dark matter halo formation}

\author[L. Lucie-Smith et al.]{
Luisa Lucie-Smith,$^{1}$\thanks{E-mail: luisa.lucie-smith.15@ucl.ac.uk}
Hiranya V. Peiris,$^{1,2}$
Andrew Pontzen,$^{1}$
\\
$^{1}$Department of Physics \& Astronomy, University College London, Gower Street, London WC1E 6BT, UK\\
$^{2}$The Oskar Klein Centre for Cosmoparticle Physics, Department of Physics, Stockholm University, AlbaNova, Stockholm, SE-106 91, Sweden
}

\begin{document}
\label{firstpage}
\pagerange{\pageref{firstpage}--\pageref{lastpage}}
\maketitle

\begin{abstract}
We present a generalization of our recently proposed machine learning framework, aiming to provide new physical insights into dark matter halo formation. We investigate the impact of the initial density and tidal shear fields on the formation of haloes over the mass range $11.4 \leq \log(M/M_{\odot}) \leq 13.4$. The algorithm is trained on an N-body simulation to infer the final mass of the halo to which each dark matter particle will later belong. We then quantify the difference in the predictive accuracy between machine learning models using a metric based on the Kullback-Leibler divergence. We first train the algorithm with information about the density contrast in the particles' local environment. The addition of tidal shear information does not yield an improved halo collapse model over one based on density information alone; the difference in their predictive performance is consistent with the statistical uncertainty of the density-only based model. This result is confirmed as we verify the ability of the initial conditions-to-halo mass mapping learnt from one simulation to generalize to independent simulations. Our work illustrates the broader potential of developing interpretable machine learning frameworks to gain physical understanding of non-linear large-scale structure formation.
\end{abstract}

\begin{keywords}
large-scale structure of Universe -- galaxies: haloes -- methods: statistical -- dark matter
\end{keywords}



\section{Introduction}
In the $\Lambda$-Cold Dark Matter paradigm of cosmological structure formation, galaxy formation proceeds within the potential wells of extended haloes of dark matter. The assembly history and internal properties of the dark matter haloes directly impact the later growth of the galaxies within their cores. An improved theoretical understanding of dark matter halo formation is therefore essential not only for constraining the fundamental properties of dark matter but also for understanding the intricate connection between halo and galaxy formation.

The evolution of dark matter haloes is determined by a series of complex, non-linear physical processes involving smooth mass accretion and violent mergers with smaller structures. For decades, N-body simulations have been used to model the non-linear evolution of haloes (e.g. \citealt{Springel2005}). 
Alongside these, simpler approximate analytic models of halo collapse can provide qualitative understanding of the results of numerical simulations. For example, extended Press-Schechter (EPS) theory and Sheth-Tormen (ST) theory are two widely accepted analytic frameworks used to infer statistical properties of dark matter haloes starting from an initial Gaussian random field. EPS theory is based on the assumption that halo collapse occurs spherically, once the smoothed linear density contrast exceeds a certain threshold \citep{PS, Bond}. The ST formalism is an extension of EPS theory to an ellipsoidal collapse model which accounts for the effect of tidal shear forces around initial peaks \citep{Doroshkevich, Bond&Myers}. These models require restrictive assumptions about the physical process of halo collapse; the haloes' non-linear evolution is approximated as spherical and ellipsoidal respectively, and formulated using excursion set theory.

Machine learning provides a tool that is well suited to modelling cosmological structure formation, given its ability to learn non-linear relationships. In fact, machine learning tools have already proved useful in the context of structure formation in, for example, distinguishing between cosmological models \citep{Merten2019} or constructing mock dark matter halo catalogues \citep{Berger2019}. However, understanding the inner workings of machine learning models remains a challenge. Developing tools to turn ``black-box'' algorithms into interpretable ones is essential for machine learning applications to physics problems; it will allow us to interpret results in terms of the underlying physics.

In \citet{LucieSmith2018}, we proposed a machine learning approach which aims to provide new physical insights into the physics of the early universe responsible for halo collapse. A machine learning algorithm is trained to learn the relationship between the early universe and late-time haloes from N-body simulations. Unlike existing analytic theories, our machine learning approach does not require modelling halo collapse with an excursion set theory; the haloes' non-linear dynamics is learnt directly from N-body simulations. The algorithm's learning is based on properties of the linear initial condition fields surrounding each dark matter particle. Machine learning algorithms are sufficiently flexible to include a wide range of properties of the initial conditions which may contain relevant information about halo formation, without changing the training process of the algorithm. By comparing the predictive performance of the algorithm when provided with different types of inputs, one can gain insights into which aspects of the early universe impact the later formation of dark matter haloes.

In \citet{LucieSmith2018}, we focused on the simplest case of a binary classification problem; the algorithm classified dark matter particles into two classes, depending on whether or not they will form haloes above a specified mass threshold at $z=0$. Contrary to existing interpretations of the Sheth-Tormen ellipsoidal collapse model, we found that the tidal shear field does not contain additional information over that contained in the density field about whether haloes will form above and below a mass threshold $M_{\mathrm{th}} = 1.8 \times 10^{12}~\mathrm{M_{\odot}}$. However, these conclusions were limited to this single mass threshold.

The aim of this work is to extend our machine learning framework to investigate haloes across a wider range of final mass. In practice, we train a machine learning algorithm to predict the value of the final mass of the halo to which each particle will belong. This is now a regression problem since the algorithm's prediction consists of a continuous variable, rather than a class label. We compare the halo mass predictions resulting from two machine learning models, trained on different sets of inputs: one on information about the initial linear density field only, and the other on both density and tidal shear information. The inputs to the algorithm, known as \textit{features} in machine learning terms, are the same as those adopted in \citet{LucieSmith2018}. We are able to quantify the relevance of the information contained in the tidal shear relative to that in the density field by comparing the predictions resulting from one model with the other. In this work, we mainly focus on the formation of haloes at $z=0$, but also verify that our conclusions hold for haloes at higher redshifts.

The paper is organized as follows. We describe the method in Sec.~\ref{sec:method}, starting with an overview of the pipeline. We then introduce the machine learning algorithm adopted in this work and describe its training and testing procedure. We present the halo mass predictions in Sec.~\ref{sec:predictions}, including a study of the algorithm's performance as a function of halo properties. We introduce a metric to make a quantitative comparison of machine learning models in Sec.~\ref{sec:metriccomparison}. We further test the generality of our results on independent simulations in Sec.~\ref{sec:generalisation}, and finally conclude in Sec.~\ref{sec:conclusions}.

\section{Method}
\label{sec:method}
In this paper we made use of six dark-matter-only simulations produced with \texttt{P-GADGET-3} \citep{gadget2, gadget} and a WMAP$5$ $\Lambda$CDM cosmological model\footnote{The cosmological parameters are $\Omega_{\Lambda} = 0.721$, $\Omega_{\mathrm{m}} = 0.279$, $\Omega_{\mathrm{b}} = 0.045$, $\sigma_{8} = 0.817$, $h = 0.701$ and $n_s = 0.96$.} \citep{WMAP}. Adopting an updated set of cosmological parameters (e.g. from \citealt{Planck2018}) is not necessary for the purpose of this work. We call the simulations \textit{sim-\#}, where $\# \in [1, 6]$. Each simulation is based on a different realization of a Gaussian random field drawn from the initial power spectrum of density fluctuations. All simulations consist of a box of comoving size $L = \num{50} \ h^{-1} \si{Mpc}$ and $N=256^3$ dark-matter particles evolving from $z=99$ to $z=0$.\footnote{We made use of the Python package \texttt{pynbody} \citep{pynbody} to analyse the information contained in the simulation snapshots.}

Dark matter haloes were identified at $z=0$ using the \texttt{SUBFIND} halo finder \citep{gadget}, a friends-of-friends method with a linking length of $0.2$, with the additional requirement that particles in a halo be gravitationally bound. While \texttt{SUBFIND} also identifies substructure within haloes, we considered the entire set of bound particles that make up a halo and did not subdivide them further. The resolution and volume of the simulation limit the resulting range of halo masses; the lowest mass halo has $M=2.6\times10^{10}~\mathrm{M}_{\odot}$ and the highest mass one $M=4.1\times10^{14}~\mathrm{M}_{\odot}$.

To train and test the machine learning algorithm, we first established the link between the initial and final state of each dark matter particle in the simulations. We used the final snapshots ($z=0$) to label each particle with the logarithmic mass of the halo to which that particle belongs. Particles that do not collapse into haloes make up $\sim50\%$ of all particles in the simulations, implying a strong class imbalance between particles not in resolved haloes and those spread across haloes of different mass scales. Training the algorithm to learn such an imbalanced mapping strongly degraded the accuracy of the predictions for particles within resolved haloes. Since our goal is to derive insight into resolved physics, we chose to restrict our analysis to the subset of particles that collapse into resolved haloes at $z=0$. Out of these, each particle, with its logarithmic halo mass label, was then traced back to the initial conditions where we extracted features to be used as input to the machine learning algorithm. 

The algorithm was trained and tested independently on the six different simulations. This yielded six different machine learning models of the same underlying mapping, allowing us to estimate the statistical significance of our results. For each simulation, the algorithm was trained based on the input features to logarithmic halo mass mapping of a training subset of particles. The remaining dark matter particles in the simulation were then used to test the algorithm's predictions against their respective true logarithmic halo mass. We will initially present the results from \textit{sim-}$1$, but we will draw the final conclusions based on the results from all six simulations.

\subsection{Gradient Boosted Trees}
\label{sec:GBT}
We used \emph{gradient boosted trees} \citep{Freund1997, Friedman2001, Friedman2002}, a machine learning algorithm combining multiple regression decision trees into a single estimator.

A regression decision tree is a model for predicting the value of a continuous target variable by following a simple set of decision rules inferred from the input features. Since individual trees generally over-fit the training data, they are often combined together to form a more robust ensemble estimator. The two main approaches to combine decision trees are \textit{bagging} and \textit{boosting}. The two approaches form ensembles that differ substantially in the trade-off between the models' ability to minimize bias and variance in the predictions. 
Bagging estimators are effective at decreasing variance, but have no effect on the bias; trees learn independently on bootstrapped training samples and the final prediction of the ensemble is given by the average over individual trees' predictions. On the other hand, boosting can reduce both the bias and the variance contributions to the error in the predictions \citep{schapire1998} by aggregating trees iteratively, such that subsequent trees learn to correct the mistakes of the previous ones. We chose to use boosting estimators, as the bias and variance of the predictions in our dataset both contribute to the predictive error.

We adopted gradient boosted trees, where the learning proceeds as follows. At any given iteration $m$ in the gradient boosted tree, a new decision tree $f_{m} (\mathbf{x})$ is added to the existing ensemble $F_{m-1}(\mathbf{x})$ such that the prediction for a given training sample $i$, $F_{m}(\mathbf{x}_i)$, is updated as
\begin{equation}
	F_{m}(\mathbf{x}_i)=F_{m-1}(\mathbf{x}_i)+ f_{m} (\mathbf{x}_i),
\end{equation}
where $\mathbf{x}_i$ is the input vector for that training sample. The accuracy of the gradient boosted tree is quantified by the loss function, a measure of how well the model's learnt parameters fit the data. The aim is to build a sequence of $M$ trees which minimizes the loss function between the target value $y$ and the predicted one $\hat{y} = F_M(\mathbf{x})$. Gradient boosted trees solve this minimization problem using gradient-descent optimization. The parameters of a decision tree, consisting of both the decision rules and the target variable for that tree, are chosen to point in the direction of the negative gradient of the loss function with respect to the ensemble's predictions. As an example, consider the loss function to be the mean squared error between the target value $y$ and the prediction $\hat{y}$. At iteration $m$, the loss function $L$ is given by the mean squared error between the target value $y$ and the current prediction $\hat{y} = F_{m-1}(\mathbf{x})$ for $N$ training samples,
\begin{equation}
	L(y, F_{m-1})=\sum_{i}^{N} \frac{\left(y_{i}-F_{m-1}(\mathbf{x}_i)\right)^{2}}{2}.
\end{equation}
The negative gradient of the loss function with respect to the predictive model for each training sample $i$ is given by
\begin{equation}
	r_i = -\frac{\partial L(y, F_{m-1})}{\partial F_{m-1}} \Bigg|_{i}=y_{i}-F_{m-1}\left(\mathbf{x}_{i}\right).
\end{equation}
Therefore, when choosing the mean squared error as the loss function, the decision tree at iteration $m$ is trained to predict the residuals $r$ of the current predictions with respect to the true target values. This procedure is repeated until adding further trees does not yield further changes in the loss. Gradient boosted trees are flexible enough to minimize any loss function, as long as it is differentiable.

In addition to the predictive power of this algorithm, gradient boosted trees also allow for interpretability of their learning procedure. This is a common feature amongst ensembles of decision trees. We made use of a metric known as \emph{feature importances} \citep{f-imp} to measure the relevance of each input feature in training the algorithm to predict the correct target variable. This is a crucial aspect of our framework; it allows us to determine which features are most informative in mapping particles to the correct final halo masses. The importance of the $j$-th feature $X_j$ from a single tree $t$ of the ensemble is given by

\begin{equation}
	\mathrm{Imp}_{t}\left(X_j \right)=\sum_{n} \frac{N_{n}}{N_{t}}\left[p -\frac{N_{n_{R}}}{N_{n}} p_{R}-\frac{N_{n_{L}}}{N_{n}} p_{L}\right]
\end{equation}
where $N_t$, $N_n$, $N_{n_{R}}$, $N_{n_{L}}$ are the total number of samples in the tree $t$, at the node $n$, at the right-child node $n_{R}$ and at the left-child node $n_{L}$, respectively. The sum in the equation is over all $n$ nodes where the feature $X_j$ makes the split. The impurity $p$ is given by the choice of splitting criterion, which in our case is the mean squared error. 
The final importance of feature $X_j$ given by the ensemble of $T$ trees is the normalized sum over the importances from all trees,
\begin{equation}
	\mathrm{Imp}(X_j)=\frac{\sum_{t=1}^{T} \mathrm{Imp}_{t}\left(X_j\right)}{\sum_{j}^J{\mathrm{Imp}\left(X_j\right)}}.
\label{importance}
\end{equation}

We used the \texttt{LightGBM} \citep{LightGBM} implementation of gradient boosted trees released by Microsoft.

\subsection{Machine learning Features}
\label{sec:features}
A \emph{feature extraction} step is required amongst most machine learning algorithms, including gradient boosted trees, to extract key properties of the dark matter particles and use them as input to the algorithm. 
Following \citet{LucieSmith2018}, we used two properties of the linear density field in the local environment around dark matter particles: the overdensity and the tidal shear computed within spheres of different mass scales centred at each dark matter particle's initial position. These choices were motivated by existing analytic frameworks which provide models to predict the final mass of a halo based on similar properties of the linear density field. EPS theory argues that a spherical patch will collapse to form a halo at redshift $z$ if its average linear density contrast $\delta_L(z)$ exceeds a critical value $\delta_c(z)$, hence motivating our choice of spherical overdensities. The final mass of the halo corresponds to the matter enclosed in the \textit{largest} possible spherical region with density contrast $\delta_L=\delta_c$. 
The ST framework motivated our choice of tidal shear information. In their approach, the collapse time of a halo depends explicitly on the ellipticity and prolateness of the tidal shear field, as well as on spherical overdensities. Using these properties as machine learning features will allow us to compare the predictions to those from analytic theories based on the same input properties and test the interpretation of these models. 

We now briefly discuss how the machine learning features were constructed from the density and tidal shear fields, referring the reader to \citet{LucieSmith2018} for further details. We smoothed the density contrast $\delta (\textbf{x}) = \left[ \rho (\textbf{x}) - \overline{\rho}_m \right]/ \overline{\rho}_m $, where $\overline{\rho}_m$ is the mean matter density of the universe, on a smoothing scale $R$,
\begin{equation}
	\delta (\textbf{x}; R) = \int \delta \left( \textbf{x}^\prime \right) W_{\mathrm{TH}} \left( \textbf{x} - \textbf{x}^\prime; R \right) \text{d}^3 x^\prime,
	\label{smoothed_delta}
\end{equation}
where $W_{\mathrm{TH}} (\textbf{x} - \textbf{x}^\prime, R)$ is a real space top-hat window function which takes the form
\begin{equation}
	W_{\mathrm{TH}} (\textbf{x} - \textbf{x}^\prime,R) = \begin{cases}  
	\dfrac{3}{4 \pi R^3} &\text{ for } \left| \textbf{x} - \textbf{x}^\prime \right| \leq R, \\ 
	0  &\text{ for }  \left| \textbf{x} - \textbf{x}^\prime \right| >R.
	\end{cases}
\end{equation}
We repeated the smoothing for $50$ smoothing mass scales (which are related to the smoothing scales $R$ via $M_{\mathrm{smoothing}} = 4/3 \pi \overline{\rho}_m R^3$), evenly spaced in $\log M$ within the range $\num{3e10} \leq M_\mathrm{smoothing} / \mathrm{M}_{\odot} \leq \num{1e15}$.

From each smoothed density contrast field $\delta(\mathbf{x}, R)$, we computed the peculiar gravitational potential $\Phi(\textbf{x})$ via Poisson's equation $\nabla^2 \Phi = \delta$ and the tidal shear tensor,
\begin{equation} 
	T^{\alpha \beta}=\left[ \frac{\partial^{2}}{\partial x^{\alpha} \partial x^{\beta}}-\frac{1}{3} \delta^{\alpha \beta} \nabla^{2}\right] \Phi.
	\label{eq:shear}
\end{equation}
We assigned two shear features to each dark matter particle, the ellipticity $e_t$ and prolateness $p_t$, following the definition of \citet{Bond&Myers}\footnote{We use the eigenvalues of the tidal shear tensor to define the ellipticity and prolateness, rather than those of the deformation tensor like in \citet{Bond&Myers}.},
\begin{align}
	e_t &= t_1 - t_3, \\
	p_t &= 3 \left( t_1 + t_3 \right).
	\label{eq:ell_prol_features}
\end{align}
where $t_1$ and $t_3$ are two of the ordered eigenvalues of the tidal shear tensor (the third is not independent since $t_1 + t_2 + t_3 = 0$). The second term on the right hand side of Eq.~\ref{eq:shear} removes the density field from the tidal shear tensor since $\nabla^{2}\Phi = \delta$, implying minimal redundancy between the information contained in the density features and that of the shear features.

In summary, we constructed two feature sets; the $50$-dimensional \textit{density} feature set made of spherical overdensities, and the $150$-dimensional \textit{density and shear} feature set made of spherical overdensities, ellipticity and prolateness features. By comparing the predictive performance of the algorithm when trained on the two feature sets, we were able to test whether the addition of tidal shear information yields an improvement in predicting the formation of the final haloes.

\subsection{Training a gradient boosted tree}
\label{sec:training}

For training the gradient boosted trees, we randomly selected $500$,$000$ particles from those that collapse into haloes at $z=0$, each carrying its own set of features and final halo mass label. No improvement in the machine learning predictions was found as we increased the size of the training set to more than $500$,$000$ particles, implying that this was sufficient to yield a training set representative of the whole simulation. The remaining particles in the simulation were used as a test set; the gradient boosted trees were trained to predict the final mass of the halo in which each test set particle will end up. The predictions were then compared to the particles' true halo masses to assess the algorithm's performance.

Gradient boosted trees have hyperparameters which must be set prior to training, and which need to be optimized for any given machine learning problem. The main hyperparameters to optimize are the number of trees in the ensemble, a gradient regularization parameter and the maximum depth and number of leaf nodes in a tree. A popular approach for hyperparameter optimization is to grid-search over a specified subset of hyperparameters and select the optimal ones using $k$-fold cross validation \citep{kfold}.
In traditional $k$-fold cross validation, the training set is divided into $k$ equal-sized sets where $k-1$ sets are used for training and one is used as a validation set to test the algorithm's performance. The validation set returns a score based on a chosen scoring metric, which can be the mean squared error or the mean absolute error in the case of regression. The training/validation procedure is repeated $k$ times such that each time a different $k$ set is used for validation and the rest for training. Finally, the score for each set of hyperparameters is given by the average score over the $k$ validation sets. There are two main benefits in using $k$-fold cross validation. First, setting aside a subset of the training set for validation ensures that the hyperparameters of the algorithm do not overfit the training data. Second, averaging the score over $k$ validation sets also ensures that the hyperparameters do not overfit any single validation set.

A disadvantage of this implementation of the method is that training and validation sets are randomly selected subsets of the same training data. Therefore, this procedure is insensitive to noise present in the training data, as this will be shared amongst both training and validation sets. In our problem, constructing validation and training sets from the same simulation may lead to overfitting the training simulation and as a result, the learned map would fail to generalize to different simulations. 

To prevent this, we constructed validation sets from the dark matter particles of a different simulation to the one used for training. All simulations were trained using $5$ validation sets from \textit{sim}-$2$, except for \textit{sim}-$2$ which used the same number of validation sets from \textit{sim}-$1$. Each set consists of $50,000$ randomly chosen particles. The hyperparameter optimization procedure then followed the standard $5$-fold cross validation approach of choosing the set of hyperparameters best performing on the validation data.

\subsection{The test set particles}
\label{sec:testing}
In each simulation, the trained gradient boosted trees can be used to predict the final halo mass of all particles in the simulation in the test set. However, we restricted our analysis to a subset of test set particles satisfying two criteria. 

First, we found that gradient boosted trees make biased predictions when the true halo mass is near the limits of the mass range probed by the simulation. The predicted masses of particles in the lowest mass haloes are overestimated and those of particles in the highest mass haloes are underestimated. The closer the true halo mass to the hard cut-offs in mass, the larger the bias in the predicted masses. Since we did not want to base our analysis on predictions affected by algorithm-specific biases, we imposed a criterion to exclude dark matter particles whose predictions are dominated by this bias. 

The second criterion excludes all particles that belong to the few haloes found in the simulation at the high mass end. The reason for this will become apparent in Sec.~\ref{sec:metriccomparison}, when we compare the predicted and true number of particles within bins of halo mass. At the high mass end, there are only a few haloes and therefore a few discrete masses in the training set. Therefore, we adopted a second criterion that excludes particles with an associated mass label in the range where the shot noise in the expected number of haloes within bins of logarithmic mass is higher than a given threshold.

In practice, these two criteria were implemented as follows. Let us denote $M_\mathrm{predicted}^i$ and $M_\mathrm{true}^i$ as the predicted and true halo mass of the $i$-th particle, respectively. We split the true halo masses of all test particles in $k$ evenly-spaced intervals of logarithmic mass. In each bin, we computed the bias $b_k$ and variance $\sigma^2_k$ defined as
\begin{align}
&b_k = \frac{1}{J_k} \sum_{j=1}^{J_k} \left[ M_{k, \mathrm{predicted}}^j - M_{k, \mathrm{true}}^j \right], \label{eq:bias} \\
&\sigma^2_k = \frac{1}{J_k} \sum_{j=1}^{J_k} \Big| M_{k, \mathrm{predicted}}^j -  \overline{M}^j_{k, \mathrm{predicted}} \Big|^{2}. \label{eq:var}
\end{align}
where $M_{k, \mathrm{predicted}}^j$ and $M_{k, \mathrm{true}}^j$ are the predicted and true halo mass of particle $j$, $\overline{M}^j_{k, \mathrm{predicted}}$ is the mean of the predicted halo masses and $J_k$ is the total number of particles in the $k$-th bin. This yielded our first criterion; we excluded from the analysis all particles in bins where $b_k^2 \geq \sigma^2_k$. 

For the second criterion, we first computed the expected number of haloes in each mass bin $k$, $N_k$,
\begin{equation}
N_k = V\int_{M_{k}}^{M_{k+1}}{\frac{dn}{dM^{\prime}} \, dM^{\prime}}
\end{equation}
where $V$ is the volume of the box and $\frac{dn}{dM}$ is the number of haloes of mass $M$ per unit volume per unit interval in $M$. The latter can be parametrized by the universal functional form
\begin{equation}
\frac{dn}{dM} = f(\sigma)\frac{\overline{\rho}_m}{M} \frac{d \ln \sigma^{-1}}{dM},
\end{equation}
where $\overline{\rho}_m$ is the cosmic mean matter density and $\sigma^2(M)$ is the mass variance of the linear density field smoothed with a top-hat window function on scale $R(M)$. We adopted the function $f(\sigma)$ predicted by \citet{ShethTormen1999} as it provides a good enough approximation of our simulation's mass function at $z=0$ and is given by
\begin{equation}
f(\sigma) = A \sqrt{\frac{2a}{\pi}} \left[ 1 + \left( \frac{\sigma^2}{a \delta_{\mathrm{c}}^2} \right)^p \right] \frac{\delta_c}{\sigma} \exp \left[ - \frac{a \delta_c^2}{2 \sigma^2}\right],
\end{equation}
where $A=0.3222$, $a=0.707$, $p=0.3$ and $\delta_c = 1.686$. Finally, our second criterion imposed that all particles with halo mass label within mass bins where the expected Poisson noise in $N_k$ exceeds $30 \%$ i.e., $1/\sqrt{N_k} > 0.3$, were excluded from the analysis.

\begin{figure}
	\includegraphics[width=\columnwidth]{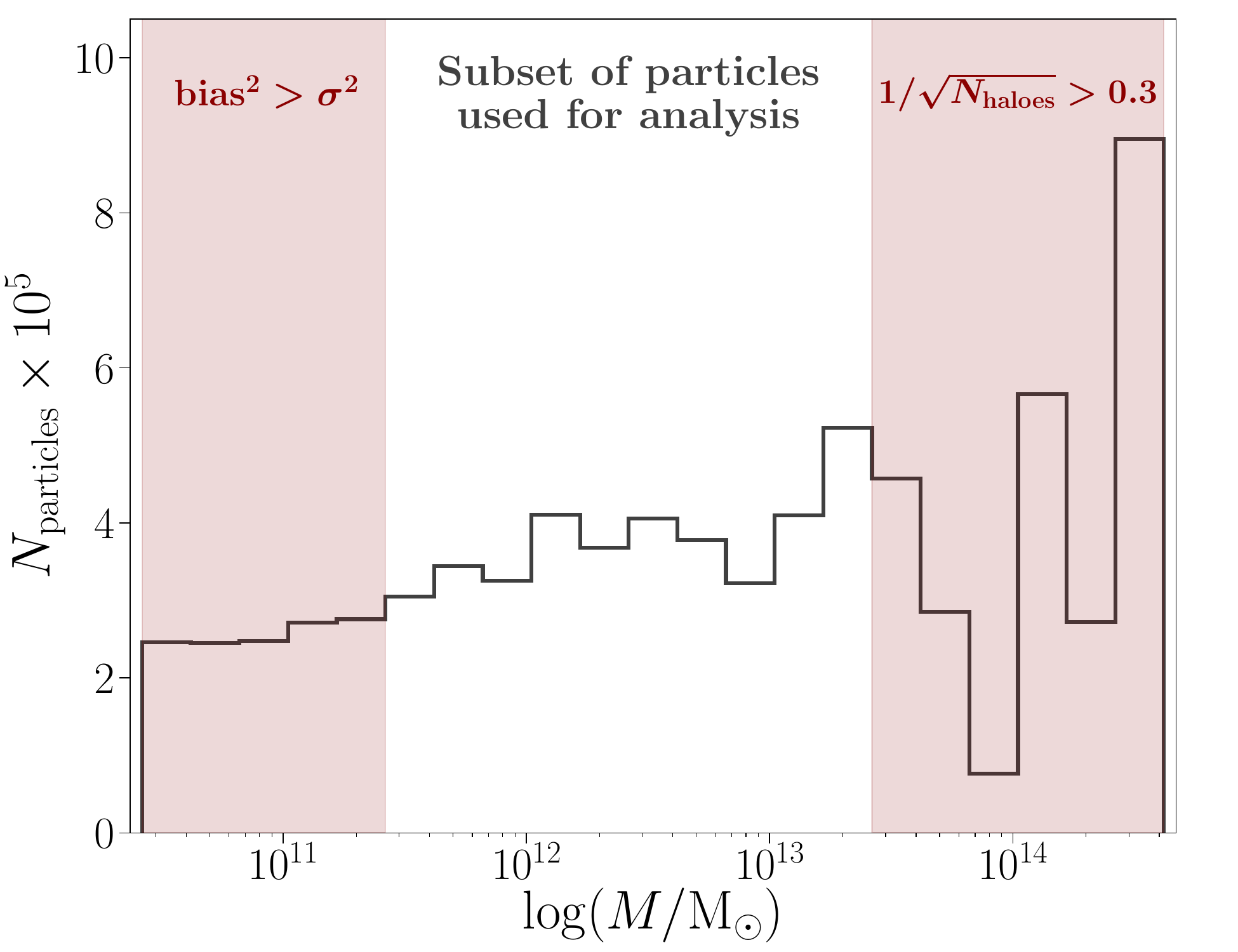}
    \caption{All particles in haloes, which were not used for training, were split into $k$ halo mass intervals of width $\Delta \log (M/\mathrm{M_{\odot}}) = 0.2$. We excluded from the analysis particles within the $k$-th mass bins where either of the following criteria are satisfied: (1) the bias in the predictions exceeds the variance i.e., $b_k^2 > \sigma_k^2$, (2) the theoretical number of haloes is smaller than a given threshold i.e., $1/\sqrt{N_{k, \mathrm{haloes}}} > 0.3$. Criterion (1) is set to exclude particles in mass bins near the mass limits imposed by the simulation, where the gradient boosted tree makes biased predictions. Criterion (2) is set to exclude mass ranges with small number of haloes. As a result, the particles used for the analysis in all simulations are those in haloes in range  $11.4 \leq \log (M/M_\odot) \leq 13.4$.}
    \label{fig:log_m_truth_distr}
\end{figure}

In summary, the subset of particles from the test set which we retained for our analysis is given by those particles belonging to haloes in $k$ mass bins where the conditions $1/\sqrt{N_k} \leq 0.3$ and  $ b_k^2 < \sigma^2_k$ are simultaneously satisfied. Both criteria are subject to the choice of bin width defining the $k$ bins; we chose $\Delta \log (M/\mathrm{M_{\odot}}) = 0.2$.\footnote{The width is chosen in order to be left with at least ten logarithmic mass bins, after applying the criteria.} The criteria were applied to all simulations, for the same choice of bin width. In all simulations, this implied that we retained particles in haloes of mass within the range $11.4 \leq \log (M/M_\odot) \leq 13.4$ for our analysis. Fig.~\ref{fig:log_m_truth_distr} shows the \textit{sim}-$1$ distribution of test set particles in haloes per logarithmic mass intervals, where the shaded regions indicate the mass ranges excluded from the analysis.

\section{Halo mass predictions}
\label{sec:predictions}
\begin{figure}
	\includegraphics[width=\columnwidth]{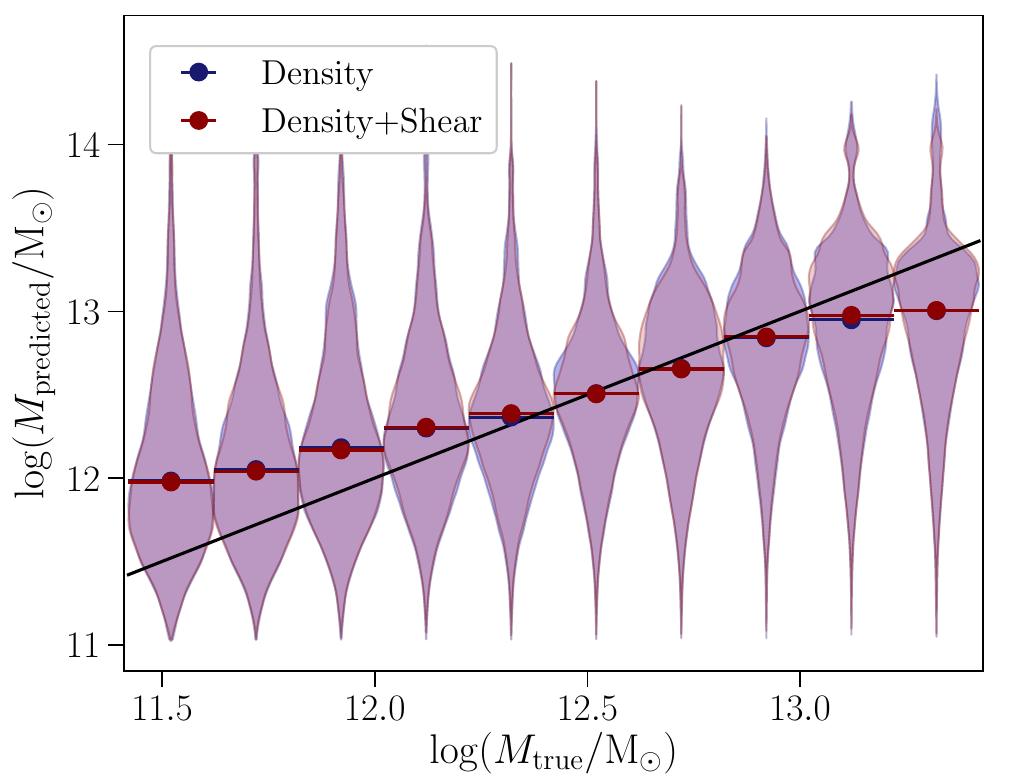}
    \caption{Distributions (and their medians) obtained with the predicted halo masses of particles within bins of width $\Delta \log (M/\mathrm{M_{\odot}}) = 0.2$, defined by their true logarithmic halo mass. The distributions are in the form of violin plots i.e., box plots whose shapes indicate the distribution of mass values. Within each bin, we compare the distributions predicted by the two machine learning models; one based on density features alone and the other based on both density and shear features. These are near-identical, meaning that there is no qualitative improvement resulting from providing the algorithm with additional information about the tidal shear field.}
    \label{fig:violins}
\end{figure}

Figure~\ref{fig:violins} compares the machine learning predictions with the true halo masses of the test set particles in \textit{sim}-$1$. 
We show the distributions obtained with the predicted halo masses of particles within bins defined by their true logarithmic halo mass. These are shown as violin plots i.e., box plots whose shapes indicate the distribution of mass values. The dots represent the medians of the predicted distributions as a function of the medians within each true mass interval. We compare the distributions resulting from two distinct machine learning models; one trained on the density feature set and the other on the density and shear feature set. We find near-to-identical predicted distributions and overlapping medians across the full mass range of haloes. We measure the fractional change in the bias and variance (as defined in Eq.~\eqref{eq:bias} \& \eqref{eq:var}) of the distributions returned by the density+shear model relative to those of the density-only model for each mass bin; we find an average change of $8\%$ in the variance and $<1\%$ in the bias. We conclude that the addition of tidal shear does not provide major qualitative changes to the predicted final mass of haloes in the range $11.4 \leq \log (M/M_\odot) \leq 13.4$, thus generalizing the conclusions of \citet{LucieSmith2018} to regression over this mass range.

\begin{figure}
	\includegraphics[width=\columnwidth]{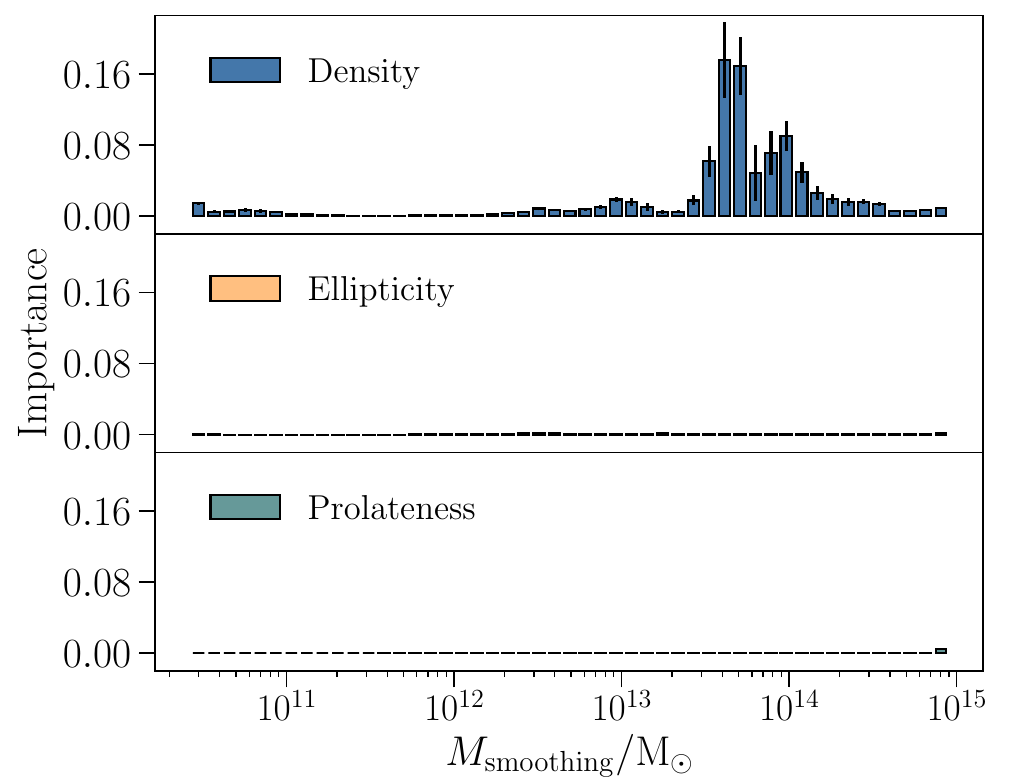}
    \caption{Feature importances for density (upper panel), ellipticity (middle panel) and prolateness (lower panel) as a function of the top-hat window function smoothing mass scale, when the gradient boosted trees are trained on the shear and density feature set. The ellipticity and prolateness features have very low importance scores, meaning that they are irrelevant compared to the density features during the training process of the algorithm. The density features are most relevant at high smoothing mass scales. This confirms that the shear field contains very little useful information compared to spherical overdensities.}
    \label{fig:imp}
\end{figure}

We now quantify which features contain the most relevant information on final halo masses, by calculating feature importances (see Sec~\ref{sec:GBT}) in the density+shear model. Fig.~\ref{fig:imp} shows that spherical overdensities on smoothing scales $10^{13} \leq M_\mathrm{smoothing}/ \mathrm{M}_{\odot} \leq 10^{14}$ are most informative for predicting the mass of haloes in the range $11.4 \leq \log (M/M_\odot) \leq 13.4$. The importances of the density features in the density-only model also have a peak and a spread at similar smoothing mass scales. The low importance of the shear features indicates that these have very little impact on the training process of the algorithm. This confirms that information about the tidal shear is not useful compared to that of spherical overdensities. 

\begin{figure}
	\includegraphics[width=\columnwidth]{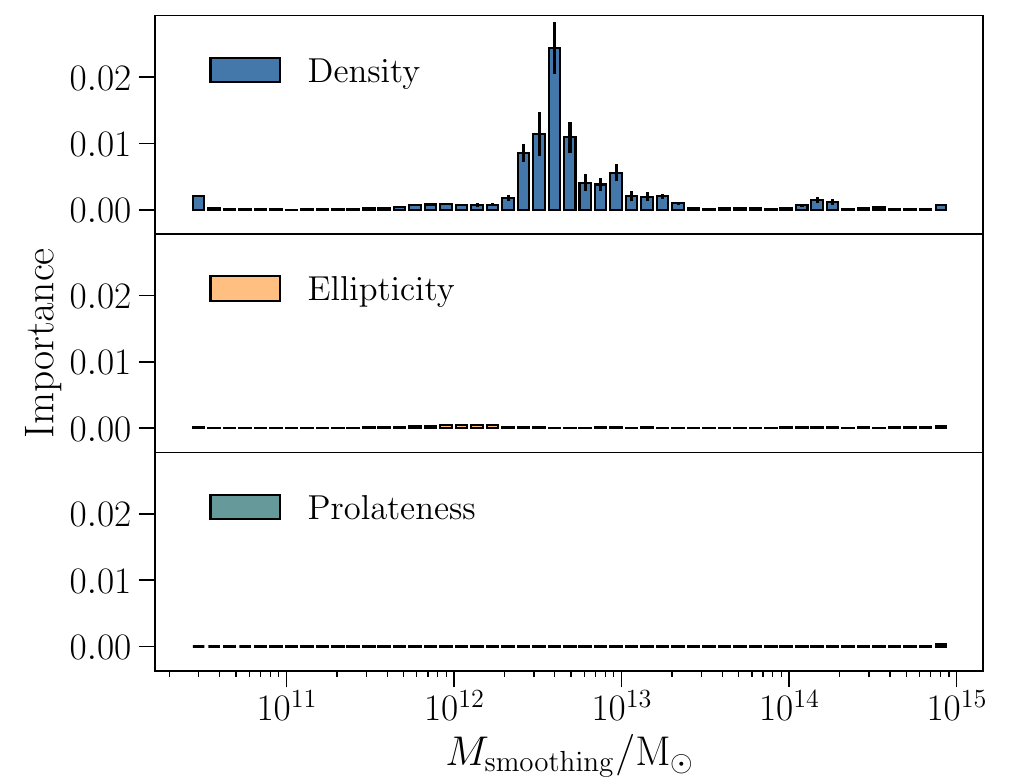}
    \caption{Feature importances for density (upper panel), ellipticity (middle panel) and prolateness (lower panel) as a function of the top-hat window function smoothing mass scale, for the case where the algorithm is trained to predict the mass of the halo to which each dark matter particle will belong at $z=2.1$. Similar to the $z=0$ case, the ellipticity and prolateness features have very little impact on the training process of the algorithm and the most relevant information is contained within the density features. The peak of the density feature importances shifts towards smaller smoothing scales, as a result of larger scales still being in the linear regime at $z=2.1$.}
    \label{fig:imp_z=2.1}
\end{figure}

We now show that this result also holds when the algorithm is trained to infer the formation of haloes at higher redshifts. Fig.~\ref{fig:imp_z=2.1} shows the density and shear feature importances, for the case of training the algorithm to predict the mass of the halo to which each dark matter particle will belong at $z=2.1$. Similar to the $z=0$ case, the ellipticity and prolateness features have negligible importance scores, meaning that the tidal shear field contains no additional relevant information over that contained in the density features about the formation of haloes at early times. The density feature importances peak at smaller smoothing mass scales, i.e. $10^{12} \lesssim M_\mathrm{smoothing}/ \mathrm{M}_{\odot} \lesssim 10^{13}$, directly reflecting the fact that larger scales are still linear at $z=2.1$ and consequently, haloes of mass $M \gtrsim 4 \times 10^{13}~\mathrm{M_\odot}$ have not yet formed.

To ensure our results capture at least as much information in the features as existing approximations, we validate the $z=0$ machine learning models against existing analytic approximations. We compare the accuracy of the machine learning predictions against those of analytic theories which also provide final halo mass predictions based on the same initial conditions information. We expect the machine learning algorithm to perform (at least) as well as analytic models. If this was not the case, it would indicate that the features contain relevant information which the algorithm fails to learn, which would in turn invalidate our conclusions. The results are shown in Appendix~\ref{sec:PS_ST_theory}; analytic and machine learning based models yield qualitatively comparable predictions, but with smaller scatter in the predictions of the machine learning model.

\subsection{Dependence on radial positions}

\begin{figure*}
	\includegraphics[width=0.98\textwidth]{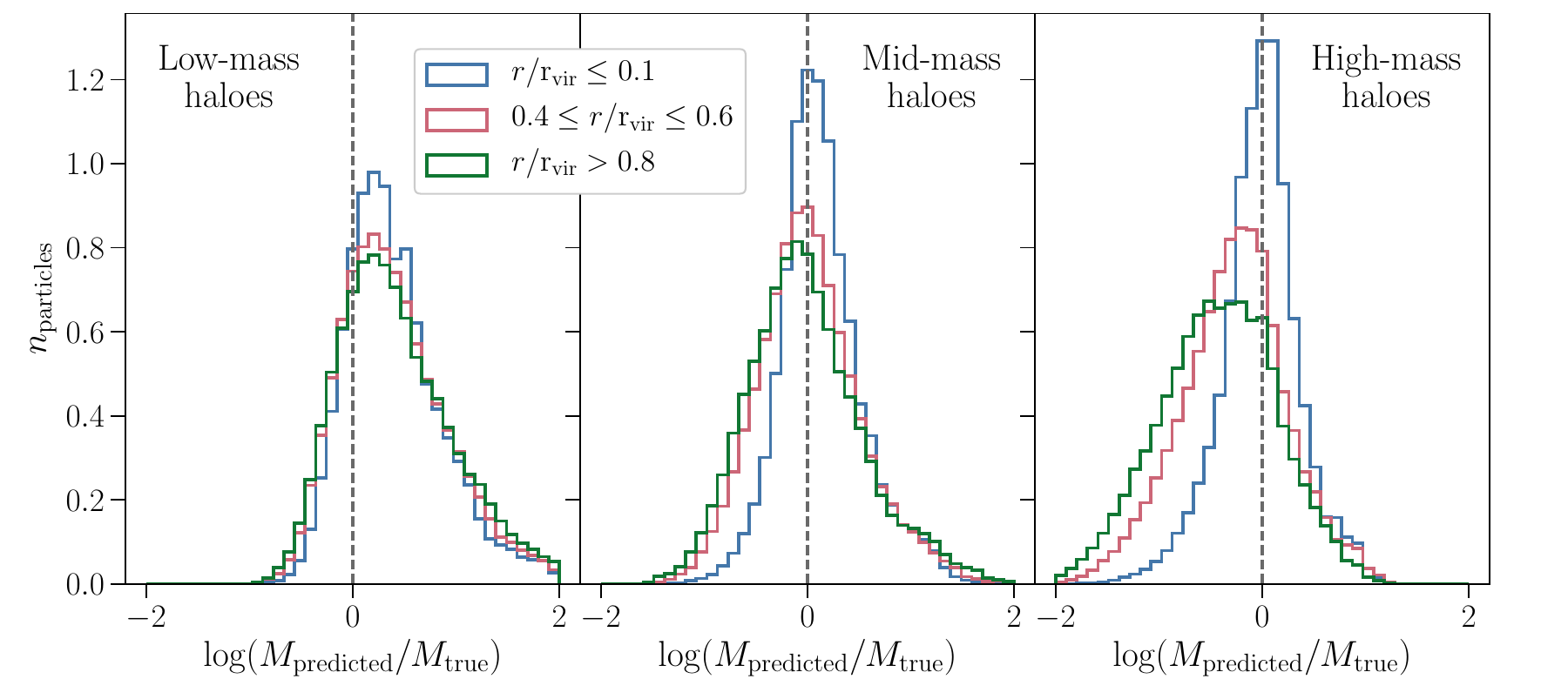}
    \caption{Distributions of $\log(M_\mathrm{predicted}/M_\mathrm{true})$ values for particles of different categories based on their radial position inside haloes. The panels show the distributions for particles in low-mass haloes  (\textit{left}), $11.42 \leq \log(M/\mathrm{M_{\odot}}) < 12.08$,  mid-mass haloes (\textit{center}) ,$12.08 \leq \log(M/\mathrm{M_{\odot}} < 12.75$, and high-mass (\textit{right}) haloes, $12.75 \leq \log(M/\mathrm{M_\odot} \leq 13.4$. The predictions of particles in low-mass haloes are uncorrelated with the particles' radial position inside the halo. For mid-mass and high-mass haloes, particles in the innermost regions of haloes are those with highest accuracy in their predicted halo masses, compared to mid-radial and outskirts particles.}
    \label{fig:rad}
\end{figure*}

We next investigated the dependence of the predictions on the radial position of particles inside haloes. This analysis was done separately for three different mass ranges of haloes. We first sub-divided particles into three equally-spaced mass ranges based on the mass of their host halo: particles in low-mass haloes ($11.42 \leq \log(M/\mathrm{M_{\odot}}) < 12.08$),  particles in mid-mass haloes ($12.08 \leq \log(M/\mathrm{M_{\odot}} < 12.75$) and particles in high-mass haloes ($12.75 \leq \log(M/\mathrm{M_\odot} \leq 13.4$). For each halo mass range, we further split the particles into three categories based on their radial position with respect to the halo's virial radius $r_\mathrm{vir}$: particles in the innermost region of a halo ($r/r_\mathrm{vir} \leq 0.1$), those in a shell of mid radial range ($0.4 \leq r/r_\mathrm{vir} \leq 0.6$) and those in the outskirts of haloes ($r/r_\mathrm{vir} > 0.8$). 

Figure~\ref{fig:rad} shows the distributions of $\log(M_\mathrm{predicted}/M_\mathrm{true})$ values of particles in each radial category predicted by the machine learning algorithm based on the density features. The three panels show the predictions of particles in low-mass (left), mid-mass (center) and high-mass (right) haloes. 
For low-mass haloes, the comparison between the distributions of the three radial categories shows very little difference, indicating that the machine learning algorithm predicts the final halo mass irrespective of their final position inside the haloes. On the other hand, we find a clear improvement in the predictions for particles in the innermost regions of mid-mass and high-mass haloes. The variance of the inner particles' predictions decreases by $35\%$ and $45\%$ for mid-mass and high-mass haloes respectively, compared to the variance of the mid-radial particles' predictions. In high mass haloes, we also note a reduction in the bias of the distributions as one approaches the haloes' central region; the medians of the $\log(M_\mathrm{predicted}/M_\mathrm{true})$ distributions are $-0.0006$, $-0.2527$ and $ -0.4101$ for inner, mid and outer radial categories, respectively. The density and shear model produces similar distributions to those returned by the density-only model.

The correlation between the accuracy of the predictions and the radial positions of particles inside their haloes is present in high mass haloes but not within low-mass ones. One possible reason for this may be the inherent difference in their assembly history. Low-mass haloes tend to accrete most of their mass at early times, whilst more massive haloes show substantial late-time mass accumulation \citep{Wechlser}. As high mass haloes are thought to undergo a larger number of merger events  \citep{Genel2008, Fakhouri2010}, the haloes may be characterized by a more complicated assembly history. In particular, particles in the outskirts of these haloes will be those that are particularly affected by late-time mergers, thus making it more difficult for the machine learning algorithm to infer their final halo mass based on their initial state.

\section{A metric for machine learning model comparison}
\label{sec:metriccomparison}

Up to this point, we have made conclusions based on visual comparisons between the predictions based on the density feature set and the density and shear one. Qualitatively, we find that the addition of tidal shear information does not yield major changes in the halo mass predictions across the whole mass range considered here. However, we require a quantitative measure of the comparison to assert whether the tidal shear contains any information that allows for a better description of halo collapse, even if minimal. 

To our knowledge, there exists no metric used in machine learning regression problems suitable for judging whether one machine learning model is preferred over another. Some of the most popular metrics used to quantify the quality of the predictions are the mean absolute error, the mean squared error or the coefficient of determination ($r^2$). These are summary statistics which provide a measure of the magnitude of the predictive error, but have no principled statistical basis and are therefore not helpful for model comparison. As one cannot construct a likelihood function from a single generative model for making predictions, we seek a metric which is (i) based on a motivated statistic and (ii) independent from the loss function optimized by the algorithm during training.

We now describe the construction of a metric which allows us to evaluate and compare the performance of machine learning models based on different feature sets. Given a set of particles and their associated halo mass labels, one can compute the number density of particles in haloes as a function of halo mass. Although the number density of particles is directly related to the number density of haloes, the resulting halo mass function cannot be meaningfully compared to existing theoretical halo mass functions due to the small range of halo masses probed by our simulations. Therefore, we choose to work with the particle number density as it is more directly related to the machine learning predictions and to the purpose of our work. The particles' ground truth halo mass labels yield a true number density distribution, $n_{\mathrm{true}}$, and those predicted by the machine learning algorithm yield a predicted number density distribution, $n_{\mathrm{ML}}$.
By comparing the two distributions, we can assess how well the machine learning approximation matches the ground truth given by the simulation. To address this question, the performance of the algorithm can be measured in terms of a difference between two distributions. In order to quantify this, we adopt the widely used Kullback-Leibler (KL) divergence \citep{kullback1951}.

The KL divergence is a measure rooted in information theory of the difference between two probability distributions. In general, the KL divergence of distribution $Q$ from $P$, $D_{\mathrm{KL}}\infdivx{P}{Q} $, describes the loss of information when $Q$ is used to approximate the reference distribution $P$. This is not a symmetric function, as the information content in $Q$ about $P$ is not equivalent to information content in $P$ about $Q$. Since we are interested in assessing how well the machine-learnt distribution describes the true distribution in the simulation, we consider the KL divergence $D_{\mathrm{KL}}\infdivx{n_{\mathrm{true}}}{n_{\mathrm{ML}}}$. If $n_{\mathrm{true}}(\log{M})$ and $n_{\mathrm{ML}}(\log{M})$ are continuous density distributions, the KL divergence takes the form
\begin{gather}
	D_{\mathrm{KL}}\infdivx{n_{\mathrm{true}}}{n_{\mathrm{ML}}} = \int_{M_\mathrm{min}}^{M_\mathrm{max}} n_{\mathrm{true}}(\log{M}) \ln \left[ \frac{n_{\mathrm{true}}(\log{M})}{n_{\mathrm{ML}}(\log{M})} \right] \mathrm{d}\log{M},
	\label{eq:KL-general}
\end{gather}
where $M_\mathrm{min}$ and $M_\mathrm{max}$ are given by the minimum and maximum values of  $\log{M}$ where $n_\mathrm{truth} (\log{M}) \neq 0$.
It is a non-negative quantity and takes the value $D_{\mathrm{KL}}\infdivx{n_{\mathrm{true}}}{n_{\mathrm{ML}}}  = 0$ if and only if the two distributions are identical i.e., $n_{\mathrm{true}}(\log{M})=n_{\mathrm{ML}}(\log{M})$. 

The KL divergence yields a machine learning model comparison metric: given two models based on different input features, the difference in the KL divergences of each model's prediction from the ground truth is a quantitative measure of the difference in the amount of information contained in one feature set over the other about final halo mass. The difference in the KL divergence for the two models is computed for each of the six simulations, allowing us to quantify its statistical significance. Our choice of metric will capture some, but not all, differences between the predictions of different models.

\subsection{Kernel density estimation}

To compute the KL divergence in Eq.~\ref{eq:KL-general}, $n_{\mathrm{true}}(\log{M})$ and $n_{\mathrm{ML}}(\log{M})$ must be in the form of continuous probability density distributions. Given the set of true and predicted mass labels of the test set particles, we can straightforwardly obtain discrete distributions for the number density of particles in haloes within bins of logarithmic mass. 
To then turn these into continuous ones, we adopted a smoothing procedure known as \textit{kernel density estimation} (KDE, \citealt{rosenblatt1956}). A KDE is a non-parametric approach to estimate the probability density distribution from a discrete set of samples. Each data point is replaced with a kernel of a set width and the density estimator is given by the sum over all kernels.

For the case of the true number density, its kernel density estimate was computed from the set of $N$ ground truth logarithmic halo masses, $\{\log M^i_{\mathrm{true}}\}_1^N$, and is given by
\begin{equation}
n_\mathrm{true}(\log M) = \frac{1}{N} \sum_{i=1}^{N} K \left( \frac{\log M - \log M^i_{\mathrm{true}}}{b} \right),
\end{equation}
where $K$ is the kernel, which we take to be a Gaussian of the form $K(x) \propto \exp(-x^2/2)$, and $b$ is a smoothing parameter known as the bandwidth, which determines the width of the kernel. 
The bandwidth is a free parameter which strongly influences the resulting estimate. If the bandwidth is too small, the density estimate will be undersmoothed and fit too closely the small-scale structure of the simulation's distribution. If the bandwidth is too large, the density estimate will be oversmoothed meaning that it will wash out important features of the underlying structure. 

We optimized the bandwidth following a $5$-fold cross validation procedure, similar to the one used to optimize the machine learning hyperparameters (see Sec.~\ref{sec:training}). For a set of bandwidth values, the KDE was fitted on the simulation's true number density distribution and validated on the distribution of an independent simulation with a different initial conditions realization. To avoid undersmoothing, we split the range of $\log{M}$ covered by the distribution into ten sub-intervals of width $\Delta \log (M/\mathrm{M_\odot}) = 0.2$ and used different mass intervals to fit and validate the KDE; every other mass bin is used for fitting and the remaining bins for validating\footnote{Note that this is the same binwidth we adopted in Sec.~\ref{sec:testing}. This choice was made to yield at least ten mass intervals for analysis, as this is the number of bins required to carry out this bandwidth optimization procedure.}. We retained the value of the bandwidth giving the highest total log-likelihood for the validation set.
\begin{figure}
	\includegraphics[width=\columnwidth]{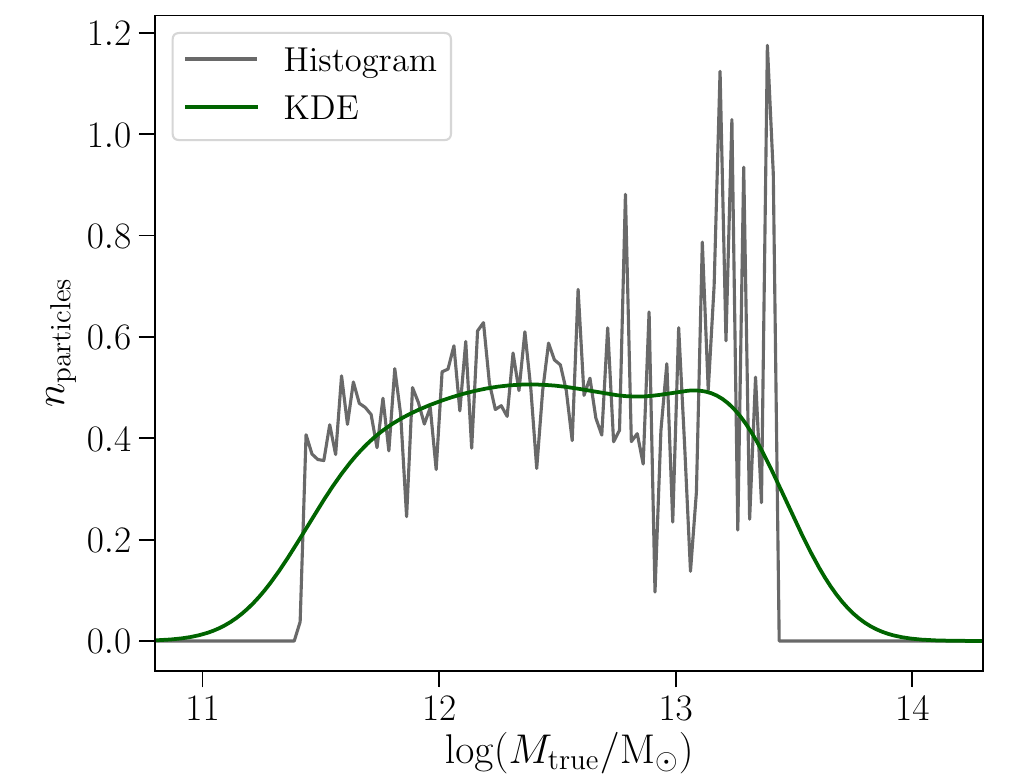}
    \caption{The distribution of test-set particles as a function of the logarithmic mass of the halo to which they belong at $z=0$. The distribution is smoothed using a kernel density estimation method, where the bandwidth is optimized using cross-validation. The upper and lower limits of the binned distribution are given by $\log(M/\mathrm{M_\odot})=11.4$ and $\log(M/\mathrm{M_\odot})=13.4$, respectively.}
    \label{fig:truth_fit}
\end{figure}

\begin{figure}
	\includegraphics[width=\columnwidth]{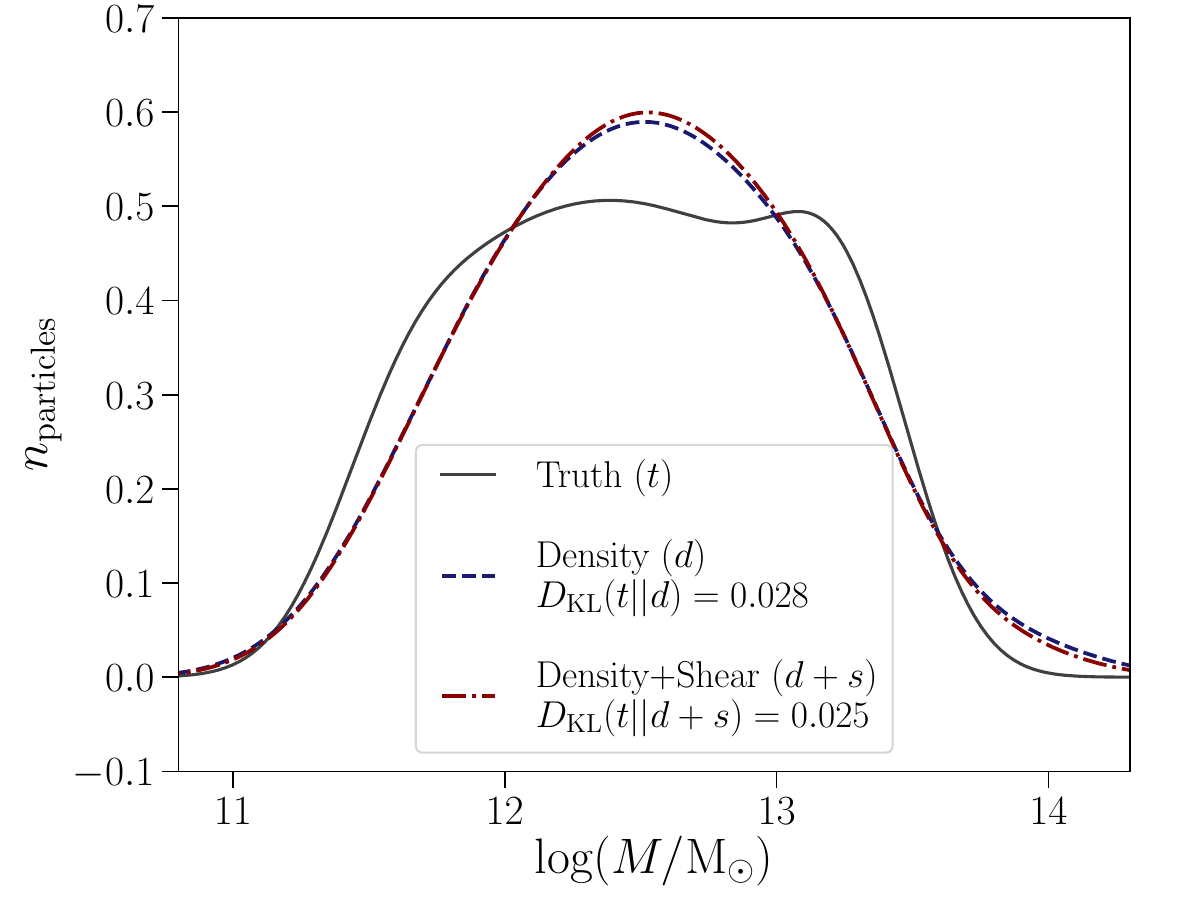}
    \caption{Predicted distribution of the \textit{sim}-$1$ test particles as a function of logarithmic halo mass for the two machine learning models, one trained with density features and the other trained on density and shear features. The ground truth distribution is also shown for comparison. We compute the KL divergence of each model's distribution with respect to the ground truth in order to quantify and compare the model's ability to approximate the true distribution. The density and shear model yields a small improvement of $0.0029$ in the KL divergence compared to the density-only model.}
    \label{fig:kde_truth_vs_ML}
\end{figure}

We smoothed each simulation's own ground truth number density of particles. For validation, all simulations used the ground truth distribution of \textit{sim}-$2$, except for \textit{sim}-$2$ which used the ground truth of \textit{sim}-$1$. All six simulations returned an optimal bandwidth $b=0.23$. The resulting kernel density estimate for \textit{sim}-$1$ is shown in Fig.~\ref{fig:truth_fit}, together with its discrete version for comparison. We then constructed density estimates from the mass values predicted by the two machine learning models, using a KDE of the same bandwidth as for the ground truth distribution. Fig.~\ref{fig:kde_truth_vs_ML} shows the comparison between the continuous number densities of particles in haloes based on the ground truth and the two machine learning models. Finally, we computed the KL divergence (as in Eq.~\ref{eq:KL-general}) for the two machine learning models with respect to the ground truth in all six simulations.

\subsection{Comparing KL divergences from different simulations}
The final step consists of comparing the KL divergences returned by the different simulations to estimate the statistical significance of our results. To do this, we first tested the validity of comparing KL divergences across different simulations. In general, a comparison between two KL divergences has a clear meaning only if they measure differences with respect to the same reference distribution. Here, the reference distributions are different; the KL divergences we wish to compare are computed with respect to each simulation's own true number density of particles in haloes. We checked whether the ground truth distributions from different realizations are similar enough for the comparison between KL divergences to be valid. We computed $D_{\mathrm{KL}}\infdivx{n_{\mathrm{true}-1}}{n_{\mathrm{true}-\#}}$, which we denote as $T$ for simplicity, to find the difference between each simulation's own ground truth distribution and that of \textit{sim}-$1$. The values of the KL divergences are reported in the last column of Table~\ref{tab:KL_table}. We find that $\overline{T}$ is at least five times smaller than any $D_{\mathrm{KL}}\infdivx{n_{\mathrm{true}}}{n_{\mathrm{ML}}}$. Therefore, the ground truth distributions are similar enough to validate the use of the KL divergence metric in the following.

\section{Results}
\label{sec:results}
\begin{table*}
	\setlength{\tabcolsep}{8pt}
	\renewcommand{\arraystretch}{1.5}
	\centering
	\caption{\label{tab:KL_table}KL divergences of a model's predicted number density of particles in haloes as a function of halo mass with respect to the ground truth distribution. Results for the density-only model ($D$) and density and shear model ($S$) of all six simulations are given in the first two numerical columns. The difference in KL divergence between the two models ($DS$)  is shown in the third column. The algorithm was trained on each simulation independently and tested on the remaining dark matter particles in that simulation not used for training. The next three columns report the KL divergences obtained with predictions made by a machine learning algorithm trained on \textit{sim}-$1$ and validated on \textit{sim}-$2$. The trained algorithm is tested on \textit{sim}-$3$, -$4$, -$5$, -$6$ and the results are shown for the density-only model ($DG$), density and shear model ($SG$) and the difference between the two ($DSG$). The last column shows the KL divergence of each simulation's own ground truth distribution and that of \textit{sim}-$1$,  $D_{\mathrm{KL}}\infdivx{n_{\mathrm{true}-1}}{n_{\mathrm{true}-\#}}$, used to validate the comparison between KL divergences of different simulations. For all columns, the last three rows show the mean, $\overline{X}$, the sample standard deviation, $\delta X$, and the standard error on the mean, $\delta\overline{X} = \delta X/\sqrt{N}$.}
	\begin{tabular}{ c|c c c | c c c | c }
		\hline
		 \textbf{Sim} & \textbf{\textit{D}} & \textbf{\textit{S}} & \textbf{\textit{DS}} & \textbf{\textit{DG}} & \textbf{\textit{SG}} &  \textbf{\textit{DSG}}&  \textbf{\textit{T}}\\ [0.5ex]
		\hline
		1 & 0.0284 & 0.0255 & 0.0029 & - & - & - & - \\
		2 & 0.043 & 0.0371 & 0.0059 & - & - & - & 0.0038 \\
		3 & 0.0419 & 0.0401 & 0.0018 & 0.0597 & 0.0616 & -0.0019 & 0.0055 \\
		4 & 0.0413 & 0.038 & 0.0032 & 0.0488 &  0.055 & -0.0062 & 0.0045 \\
		5 & 0.0387 & 0.0286 & 0.0101 & 0.0519 & 0.0577 & -0.0058 & 0.0127 \\
		6 & 0.0188 & 0.0136 & 0.0052 & 0.0361 &  0.0361 & 0 & 0.0027 \\ [0.5ex]
		\hline
		$\overline{X}$ & 0.0353 & 0.0305 & 0.0049 & 0.0491 & 0.0526 & -0.0035 & 0.0058 \\
		$\delta\overline{X}$ & 0.004 & 0.0041 & 0.0012 & 0.0049 & 0.0057 & 0.0015 & 0.0018 \\
		$\delta X$ & 0.0097 & 0.0101 & 0.003 & 0.0098 & 0.0113 & 0.003 & 0.0039 \\ [0.5ex]
		\hline
	\end{tabular}
\end{table*}

We present our results in Table~\ref{tab:KL_table}. The first three columns show the values of $D_{\mathrm{KL}}\infdivx{n_{\mathrm{true}}}{n_{\mathrm{density}}}$,  $D_{\mathrm{KL}}\infdivx{n_{\mathrm{true}}}{n_{\mathrm{density+shear}}}$ and the difference between the two, $D_{\mathrm{KL}}\infdivx{n_{\mathrm{true}}}{n_{\mathrm{density}}} - D_{\mathrm{KL}}\infdivx{n_{\mathrm{true}}}{n_{\mathrm{density+shear}}}$ for all six simulations. We call these $D$, $S$ and $DS$ respectively, to simplify the notation. 
For each column $X$, we also compute the mean over the six realizations, $\overline{X}$, the sample standard deviation, $\delta X$, and the standard error on the mean, $\delta\overline{X} = \delta X/\sqrt{N}$, where $N=6$ simulations.

The values of $DS$ indicate the change in the KL divergence as we add information about the tidal shear in all six simulations. We measured the statistical significance of the deviation of $\overline{DS}$ from $0$ given its standard error $\delta\overline{DS}$. We find an improvement in the KL divergence (at the $4$--sigma level) provided by the addition of shear information relative to a model based on density information alone. 
We quantify the practical utility of such an improvement by comparing the value of $\overline{\mathit{DS}}$ with $\delta{D}$, the scatter in the density-only model. We find that the improvement provided by shear information is equivalent to a $0.5$--sigma deviation from the mean KL divergence of the density-only model. Therefore, we conclude that the improvement provided by the tidal shear is not large enough to yield a useful alternative model to one based on density information alone. These conclusions are consistent with the results of the feature importance analysis in Sec.~\ref{sec:predictions}.

\section{A test of generalizability}
\label{sec:generalisation}
The results presented above are valid for the case where the dark matter particles that make up the training set and the test set come from the same simulation. To test the robustness of our results, we verified the ability of the machine learning algorithm trained on one simulation to generalize to independent simulations based on different initial conditions realizations. In particular, we tested whether our main results about the significance and the utility of the improvement provided by tidal shear information still holds when generalizing to independent simulations.

We used the machine learning algorithm trained on \textit{sim}-$1$ and tested it on all dark matter particles in \textit{sim}-$3$, -$4$, -$5$, -$6$, which are independent from the training process of \textit{sim}-$1$. Since the dark matter particles in \textit{sim}-2 form the validation sets used during training, we excluded the latter from this analysis. As before, we computed the KL divergences $D_{\mathrm{KL}}\infdivx{n_{\mathrm{true}}}{n_{\mathrm{density}}}$,  $D_{\mathrm{KL}}\infdivx{n_{\mathrm{true}}}{n_{\mathrm{density+shear}}}$ and the difference between the two, $D_{\mathrm{KL}}\infdivx{n_{\mathrm{true}}}{n_{\mathrm{density}}} - D_{\mathrm{KL}}\infdivx{n_{\mathrm{true}}}{n_{\mathrm{density+shear}}}$; the values of these quantities for the four independent test simulations are reported in the fourth, fifth and sixth columns of Table \ref{tab:KL_table}. This time we call these $DG$, $SG$ and $DSG$ respectively, to distinguish them from the previous case where the test set and training set are constructed from the same simulation.

First, we tested the generalisability of each machine learning model individually. For the density feature set, the mean KL divergence computed from the independent test sets ($\overline{\mathit{DG}}$) is consistent (at the $2.2$--sigma level) with that found when training and testing on the same simulation ($ \overline{\mathit{D}}$), meaning that the model learnt on one simulation can indeed generalize to independent simulations. This confirms that the machine learning algorithm was able to learn the underlying physics relating the initial conditions to the final haloes. On the other hand, the model based on density and shear features shows evidence of poor generalisability, as the KL divergences $\overline{\mathit{SG}}$ and $\overline{\mathit{S}}$ are in tension at the $3.2$--sigma level.

We then moved on to test the generalisability of our results regarding the improvement provided by the addition of tidal shear information. We find that the difference in the KL divergence of the two models ($\overline{DSG}$) is in significant tension (at the $4.3$--sigma level) with that found when testing on the same simulation used for training ($ \overline{DS}$). Moreover, as $\overline{DSG}$ was a negative value, the addition of tidal shear information now yields a marginal loss in performance, rather than an improvement. 

These discrepancies provide some evidence that the algorithm trained on density and shear features overfits the simulation during training. This naturally yields better predictions when testing the algorithm on the simulation used for training compared to testing on independent simulations. Consequently, the addition of tidal shear information yields an improvement or a loss in performance compared to the density-only model, depending on whether the algorithm is tested on the same or a different simulation to that used for training. In spite of this, the level of overfitting in the density and shear model is small; for both cases, the change in KL divergence between the two models ($\overline{DS}$, or $\overline{DSG}$) is consistent with the scatter in the density-only model ($\delta{D}$, or $\delta{DG}$). 

In summary, the algorithm trained on density information has learnt the physical connection between the initial conditions and the final haloes, as it is able to generalize to independent realizations of the initial density field. On the other hand, the improvement in the KL divergence provided by the addition of tidal shear features is lost when applying the trained algorithm to independent simulations. Therefore the improvement from including shear features in the machine learning process, which was anyway small, does not imply any physical connection. This strengthens our conclusion that there is no identifiable physical information pertinent to the final halo mass in the tidal shear field.

These conclusions were made by testing the algorithm on independent realizations with fixed cosmological parameters. 
The parameters of the $\Lambda$CDM model are so tightly constrained from current observations (e.g. \citealt{Planck2018}), that the formation of haloes must proceed in a similar way at the mass scales investigated in our analysis. Therefore, we expect no significant change in our results when adopting simulations based on different choices of cosmological parameters. Moreover, we expect similar results for the mass range considered in this analysis for observationally-allowed cosmological models which suppress small-scale power; in such models halo abundances differ from $\Lambda$CDM only below $M \sim 10^{11}~\mathrm{M_\odot}$.

Our results for the halo mass range $11.4 \leq \log(M/\mathrm{M_{\odot}}) \leq 13.4$ are also expected to hold for simulations of different box sizes or resolutions. In particular, a simulation with larger box size or higher resolution yields the possibility of extracting additional features at larger or smaller smoothing scales, respectively. Since the feature importances (Fig.~\ref{fig:imp}) show that the most relevant information is contained within features on smoothing scales $10^{13} \leq M_\mathrm{smoothing}/ \mathrm{M}_{\odot} \leq 10^{14}$, the results do not change when the simulation contains additional small- or large-scale information. Similarly, our results should hold for simulations of smaller box sizes and/or lower resolutions, as long as those scales which carry the most relevant information are resolved.

\section{Conclusions}
\label{sec:conclusions}
We have presented a generalization of the work in \citet{LucieSmith2018}, which explored the impact of different initial linear fields on the formation of dark matter haloes above or below a single mass threshold. In this paper, we investigated a wider mass range of dark matter haloes and their sensitivity to the initial density and tidal shear fields.

We find that the tidal shear field does not contain additional information over that already contained in the linear density field about the formation of dark matter haloes in the mass range $11.4 \leq \log(M/M_{\odot}) \leq 13.4$. We quantified this using a machine learning regression framework, showing that the results are physically interpretable and generalisable to independent realizations of the initial density field. Interpretability is achieved by comparing machine learning models based on different input properties of the initial conditions; the addition of tidal shear information yields a halo collapse model whose predictions are statistically consistent with those of a model based on density information alone, according to a metric based on the Kullback-Leibler divergence. By measuring the feature importances of the different inputs during the training process of the algorithm, we can establish a complementary measure of which physical aspects contain the most information about halo collapse. This analysis confirms that our machine learning approach suggests little role for the tidal shear field in establishing final halo masses. This result holds also for the case of predicting the mass of haloes at $z=2.1$. Generalisability is verified by applying the machine learning algorithm trained on one simulation to independent simulations based on different realizations of the initial density field. This allows us to confirm the ability of the machine learning algorithm to learn physical connections between the initial conditions and the final dark matter haloes.

Our work demonstrates the utility of machine learning techniques to gain physical understanding of large-scale structure formation. The strength of this approach lies in its ability to establish a physical interpretation of the machine learning results. In future work, we plan to extend our framework to develop interpretable deep learning algorithms, aiming to learn directly from the initial density field which physical aspects are most relevant to cosmological structure formation, beyond spherical overdensities and tidal shear forces.

\section*{Acknowledgements}

We thank Justin Alsing, Boris Leistedt, Michelle Lochner, Jason McEwen, Daniel Mortlock, Nikos Nikolaou, Martin Rey and Ravi Sheth for useful discussions. LLS acknowledges the hospitality of the Oskar Klein Centre, Stockholm where part of this work was completed. LLS was supported by the Science and Technology Facilities Council. HVP was partially supported by the European Research Council (ERC) under the European Community's Seventh Framework Programme (FP7/2007-2013)/ERC grant agreement number 306478- CosmicDawn, and the research project grant ``Fundamental Physics from Cosmological Surveys'' funded by the Swedish Research Council (VR) under Dnr 2017-04212. AP was supported by the Royal Society. This work was partially enabled by funding from the UCL Cosmoparticle Initiative.





\bibliographystyle{mnras}
\bibliography{mlhalos_regression} 



\appendix

\section{A comparison with analytic theories}
\label{sec:PS_ST_theory}
\begin{figure*}
	\includegraphics[width=0.8\textwidth]{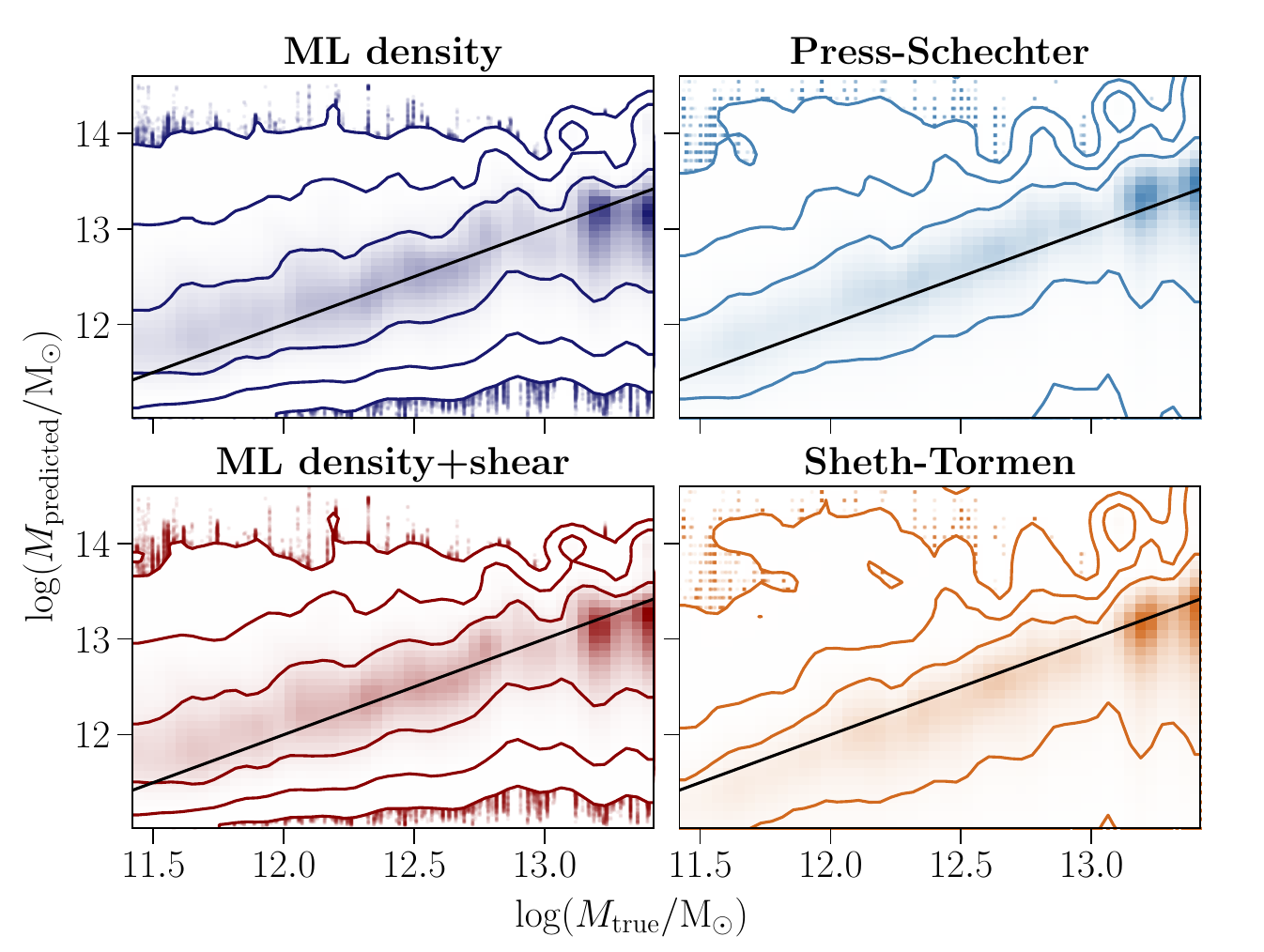}
    \caption{Two-dimensional histograms and contours containing $68\%$, $95\%$ and $99.7\%$ of the joint probability of the predicted vs. true halo masses for the analytic and machine learning models. We compare the machine learning predictions based on the density features with EPS theory and those based on density and tidal shear features with ST theory. The predictions are qualitatively similar, but with tighter confidence regions in the machine learning case. This validates our machine learning results as we find no evidence of any relevant information contained in the features that the algorithm fails to learn.}
    \label{fig:ML_vs_analytic}
\end{figure*}
We validated the machine learning findings by comparing the accuracy of its predictions against those of analytic theories which also provide final halo mass predictions based on the same initial conditions information. 

We compared the machine learning predictions based on the density features with EPS theory and those based on density and tidal shear features with ST theory, for the test set particles in \textit{sim}-$1$. According to EPS, the fraction of density trajectories with a first upcrossing of a density threshold barrier $\delta_\mathrm{th}$ is equivalent to the fraction of haloes of mass $M$. The density threshold barrier $\delta_\mathrm{th}$ adopted by \citet{Bond} is that of spherical collapse: $\delta_\mathrm{th}(z) = (D(z)/D(0))\delta_\mathrm{sc}$, where $\delta_\mathrm{sc} \approx 1.686$. The predicted halo mass of each test particle is given by the smoothing mass scale at which the particle first upcrosses the density threshold barrier. 

In the ST formalism, EPS theory is extended to account for the effect of the tidal shear field by adopting a ``moving" collapse barrier rather than the spherical collapse barrier. The ST collapse barrier $b(z)$ varies as a function of the mass variance $\sigma^2(M)$ and is given by
\begin{equation}
	b(z) = \sqrt{a} \delta_\mathrm{sc}(z) \left[ 1 + \left( \beta \dfrac{\sigma^2 (M)}{a \delta_{\mathrm{sc}}^2 (z)} \right)^{\gamma} \right], 
	\label{eq:ST_barrier}	
\end{equation}
where $\delta_{\mathrm{sc}} (0) \approx 1.686$ and the best-fit parameters found in \citet{ShethMoTormen2001} are $\beta  = 0.485$, $\gamma = 0.615$ and $a = 0.707$. Similar to the EPS case, the predicted halo mass of each test particle is given by the smoothing mass scale at which the particle first upcrosses the threshold barrier given by Eq.~\eqref{eq:ST_barrier}. In summary, for each test particles we can compute the EPS and ST predicted halo masses and compare those to the machine learning density-only and density combined with shear predictions, respectively. 

Figure~\ref{fig:ML_vs_analytic} shows the predicted halo masses as a function of true halo masses for the analytic and machine learning models. We show two-dimensional histograms and the contours containing $68\%$, $95\%$ and $99.7\%$ of the joint probability. Machine learning and analytic models show qualitatively similar predictions, but with tighter confidence regions for the machine learning predictions. This is especially notable where the analytic models' predictions extend to much lower mass values than the machine learning predictions. Note also that the ST predictions are shifted towards lower mass values compared to the PS predictions, for fixed true halo mass. This directly reflects the fact that the ST collapse barrier takes larger $\delta$ values than the PS barrier at fixed smoothing mass scale; the same particle will therefore cross the collapse barrier at lower smoothing mass scales for ST than PS, which in turn results in a lower halo mass prediction.

This test validates our machine learning results by ruling out the possibility that the algorithm is not making use of all the information contained in the features. Moreover, this also shows that the machine learning algorithm provides better predictions than the analytic models on a particle-by-particle basis.


\bsp	
\label{lastpage}
\end{document}